\documentclass{article}  
\usepackage{jamaica04}
\usepackage{graphicx}
\frompage{000} \topage{000}                                              
\def\rightmark{Landau Hydrodynamics \& RHIC Phenomenology}
\def\leftmark{Peter Steinberg}

\newcommand{\epem}{e^+e^-}

\newcommand{\np}{N_{part}}

\title{Landau Hydrodynamics and RHIC Phenomena} 
\authors{
{Peter Steinberg$^1$ %
}\\[2.812mm]
{\normalsize
\hspace*{-8pt}$^1$ Brookhaven National Laboratory, \\ 
Upton, NY 11973\\[0.2ex] 
}}
 
\abstract{
The basic physical assumptions and results of Landau's
hydrodynamic model of particle production are reviewed.
It is argued that these results have sufficient descriptive
and predictive power in strong-interaction phenomenology, 
including recent RHIC data, to warrant a closer examination
of the physical assumptions.
}

 
\begin{document}
 
\maketitle
\setcounter{page}{1}

\section{Introduction}
The success of boost-invariant hydrodynamics at RHIC (e.g. \cite{Kolb:2003dz})
to describe the systematics of
elliptic flow is considered to be strong evidence that the strongly
interacting system thermalizes early, coupling the dynamical 
evolution to initial-state geometry of the system.  
These calculations can, in principle, provide a means to extract
the form of the equation of state of the thermalized system.  This
may provide insight into whether or not a quark-gluon plasma was
in fact created in heavy ion collisions at RHIC.

It should not be forgotten, however, that hydrodynamics has a long
history in the study of strongly-interacting systems.  Coupled with
the intense interest in statistical and thermal model calculations
in the early 1950's, spearheaded by Fermi and Landau
\cite{Fermi:1950jd,Landau:gs,Belenkij:cd}, this led to
a large body of strong interaction phenomenology, manifestly 
{\it non} boost-invariant, which was refined
throughout the years
\cite{Carruthers:ws,Carruthers:dw,Carruthers:1981vs,Cooper:1974ak,Cooper:1974mv,Cooper:qi,Zhirov:qu}.
It is the goal of this talk to review the central physical
assumptions and predictions of the Landau hydrodynamical model
(and its refinements), and to show its relevance to a wide range
of results in a variety of strongly-interacting systems.

In the Landau-Fermi physical picture, the main physical assumptions were:
\begin{itemize}
\item The collision of two Lorentz-contracted hadrons or nuclei leads to
full thermalization in a volume of size $V{m_p}/\sqrt{s}$.  This
justifies the use of thermodynamics and establishes the
system size and energy dependence.
\item A massless blackbody EOS is assumed $p=\epsilon/3$.
This allows the complete calculation of physical quantities.
\item All chemical potentials (meson and baryon) are zero,
which dramatically simplifies the mathematics.
\end{itemize}
The main results derived from these assumptions are:
\begin{itemize}
\item A universal formula for the produced entropy, determined mainly
by the initial Lorentz contraction.
\item Gaussian rapidity distributions, at least for particles produced
several units away from the projectile rapidities
\item Thermal particle occupations determined by $T\sim m_{\pi}$.

\end{itemize}

\section{Universal Entropy}

\begin{figure}[t]
\begin{center}
\includegraphics[width=10cm]{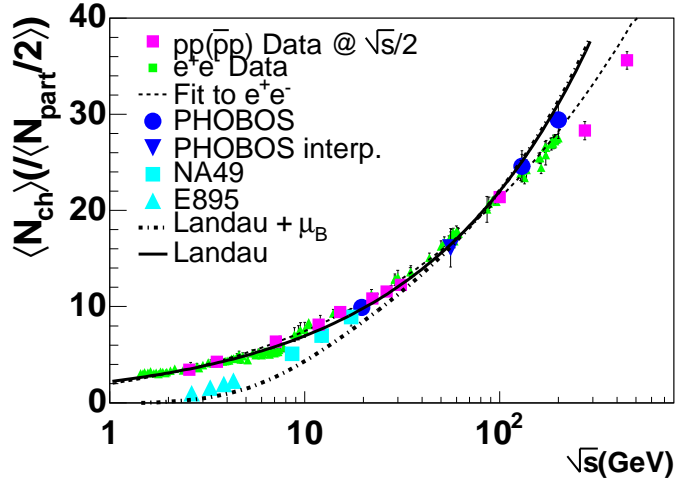}
\caption{
Charged particle multiplicities for A+A, p+p (with leading particle
effect removed) and $\epem$.  Theoretical curves are pQCD (dotted
line) \cite{Mueller:cq}, 
baryon-free Landau (solid line) \cite{Carruthers:dw}, 
and Landau including the
baryochemical potential effect (dot-dashed) \cite{cs}.
\label{s14}
}
\end{center}
\end{figure}

Simply using the first law of thermodynamics and the blackbody
EOS,
Landau and Fermi both arrived at the same scaling formula for the multiplicity
produced in a collision of two strongly interacting objects, 
$N_{ch} = \alpha S = K s^{1/4}$ \cite{Fermi:1950jd,Landau:gs,Belenkij:cd}.  
The agreement between this formula
and a wide range of systems is shown in Fig. \ref{s14} with a
constant of proportionality of K=2.2, which has been
found to work with systems as diverse as p+p, Au+Au, $\epem$
and $\nu+p$ \cite{Carruthers:ws,Carruthers:dw,Carruthers:1981vs}.

It was always controversial
to apply Landau's reasoning to ``small'' systems, such
as $\epem$ annihilations.
Beyond the usual arguments about falling far short of the thermodynamic
limit, it has been argued that perturbative QCD (pQCD) already
provides an excellent description of not just the multiplicity
of produced hadrons as a function of energy, but the details
of jet shapes (modulo hadronization corrections),
rendering a ``statistical'' description superfluous (at best!)
Despite this, it is interesting to observe that pQCD calculations
(e.g. Ref. \cite{Mueller:cq}) and the Fermi-Landau 
formula agree within 10\% over a wide range of
energies, essentially up to where the LEP data ends, as shown
in Fig. \ref{s14} (adapted from Ref. \cite{Back:2003xk}).
This similarity is striking when one considers that pQCD implies
an infinite mean-free-path for parton rescatterings, while the
hydro description implies a negligible one for the (presumably partonic) 
degrees of freedom which thermalize.

The extension of the Fermi-Landau approach from a single $p+p$
collision to nuclei is surprisingly simple.  
Since the Lorentz contraction
is not changed, the angular distributions are in principle
similar to the smaller system
\cite{Belenkij:cd}.  
Then, provided that the
interactions between the subvolumes of the system do not
themselves open up new degrees of freedom, the entropy of the system as a
whole will simply scale proportionally to the number of participating
nucleons ($N_{part}$).  This ``$\np$-scaling'' has been observed 
for the total multiplicity (but not necessarily in any particular
region of phase space) in all collisions involving nuclei, 
from $p+A$ to $Au+Au$ collisions \cite{Back:2003xk,Nouicer:2004ke}.

\section{Thermal Phenomenology and Hadrochemistry}

In the Landau scenario, freezeout is not expected to
occur immediately, as Fermi assumed, but rather when the 
temperature reaches the limit of the pion Compton wavelength
$T=m_{\pi}$.  This was based on a suggestion by Pomeranchuk \cite{Pomeranchuk:ey} 
to avoid Fermi's prediction that nucleons would outnumber pions
by virtue of their larger statistical weight.
This assumption leads to predictions for the 
relative population of various particle states
similar to those made in the Hagedorn approach 
\cite{Cleymans:2002mp,Braun-Munzinger:2003zz}.

A+A collisions clearly deviate from the Fermi-Landau formula
at low energies.  An obvious suspect is the phenomenon
of baryon stopping, which is absent in p+p collisions
but is substantial in A+A \cite{Bearden:2003hx}.
If one puts back the $-\mu_B N_B$ term into the first law of thermodynamics, we 
immediately see how the presence of a conserved quantity associated
with a substantial mass (i.e. the proton mass) will naturally
suppress the total entropy: $S = (E+pV-\mu_B N_B)/T$.
Using an existing thermal model code, Cleymans and Stankiewicz
\cite{cs}
calculated the entropy density as a function of $\sqrt{s}$.
It rises to limiting value where $\mu_B\rightarrow 0$
and $T\rightarrow T_0$, the Hagedorn temperature.  If we then assume
that the total multiplicity scales linearly with the total
entropy, then the suppression relative to the asymptotic value
should describe the energy dependence of the A+A multiplicities
relative to the Landau formula.  This allows a direct comparison
with the multiplicity data as shown in Fig. \ref{s14}.
There we find a reasonable qualitative description of the difference
between A+A and $\epem$ collisions.

\section{Gaussian Rapidity Distributions}


Up to this point, the discussions have only involved the
total entropy, integrated over the full phase space.  
Landau was apple to perform approximate calculations of the
angular distributions by introducing hydrodynamic evolution 
using the standard equations of relativistic hydrodynamics
$\partial_{\mu} T^{\mu\nu} = 0$ closed by the blackbody EOS
\cite{Landau:gs,Carruthers:dw}.
These equations generically imply that the initial state
entropy, which is produced in the process of thermalization,
is distributed in rapidity space with a
a Gaussian form, the width determined by the initial Lorentz 
contraction.  
Including the total multiplicity
formula to set the overall normalization, the full expression 
\cite{Carruthers:dw} is:
\begin{equation}
\frac{dN}{dy}=\frac{Ks^{1/4}}{\sqrt{2\pi L}}\exp(-\frac{y^2}{2L}).
\label{lanfull}
\end{equation}
where $L=\sigma^2_y=(1/2)\ln(s/m^2_p)=\ln(\gamma)$.
Already, one can see a connection between the shape of
the distribution and the multiplicity, since from this
definition $s^{1/4} \propto e^{L/2}$.


\begin{figure}[t]
\begin{center}
\begin{minipage}{60mm}
\includegraphics[width=6cm]{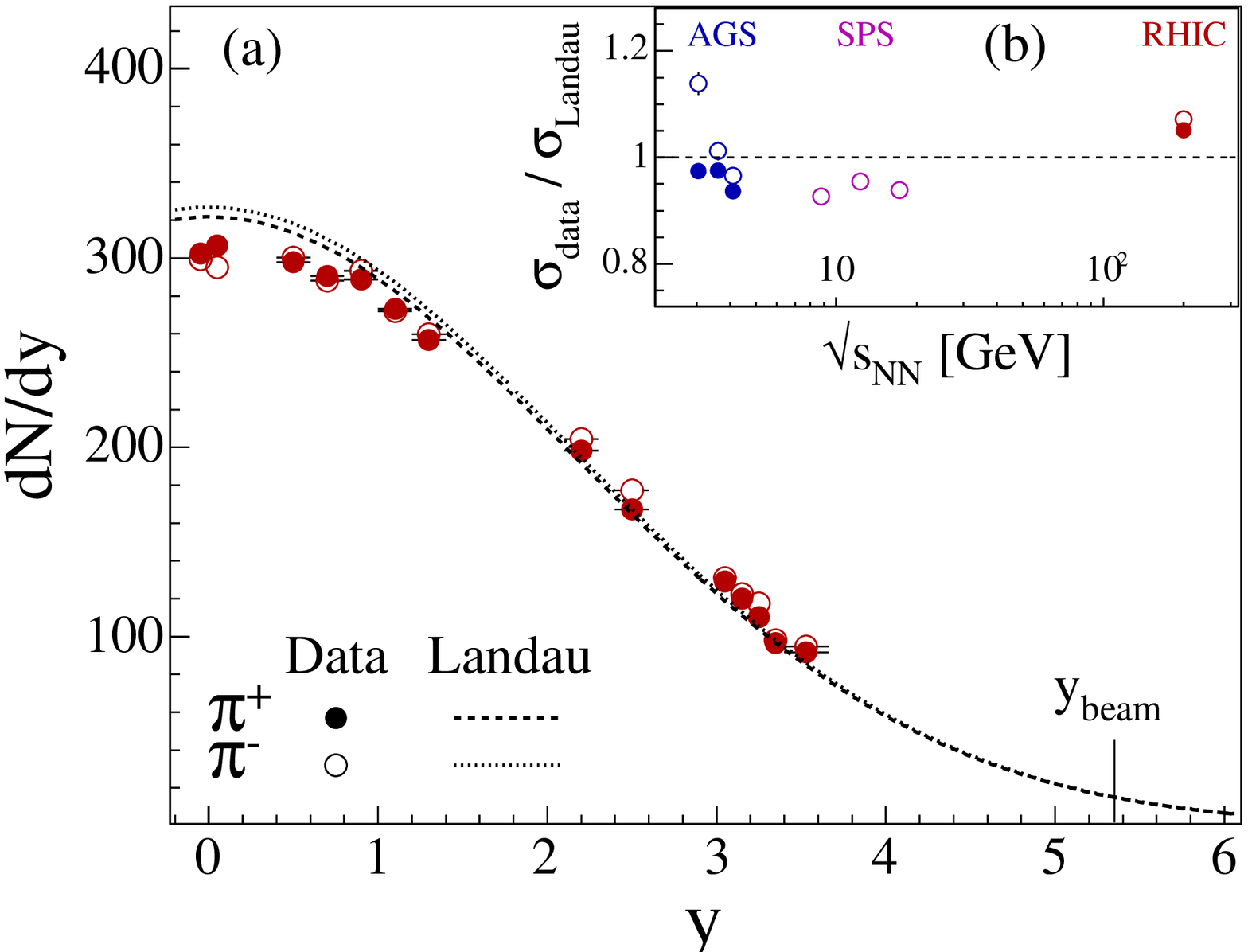}
\caption{BRAHMS data on $dN/dy$ for charged pions,
compared with Landau's prediction.  Inset is a comparison
between the measured and predicted Gaussian widths
(from Ref. \cite{Bearden:2004yx})
\label{brahms-dndy}}
\end{minipage}
\hspace{\fill}
\begin{minipage}{60mm}
\includegraphics[width=6cm]{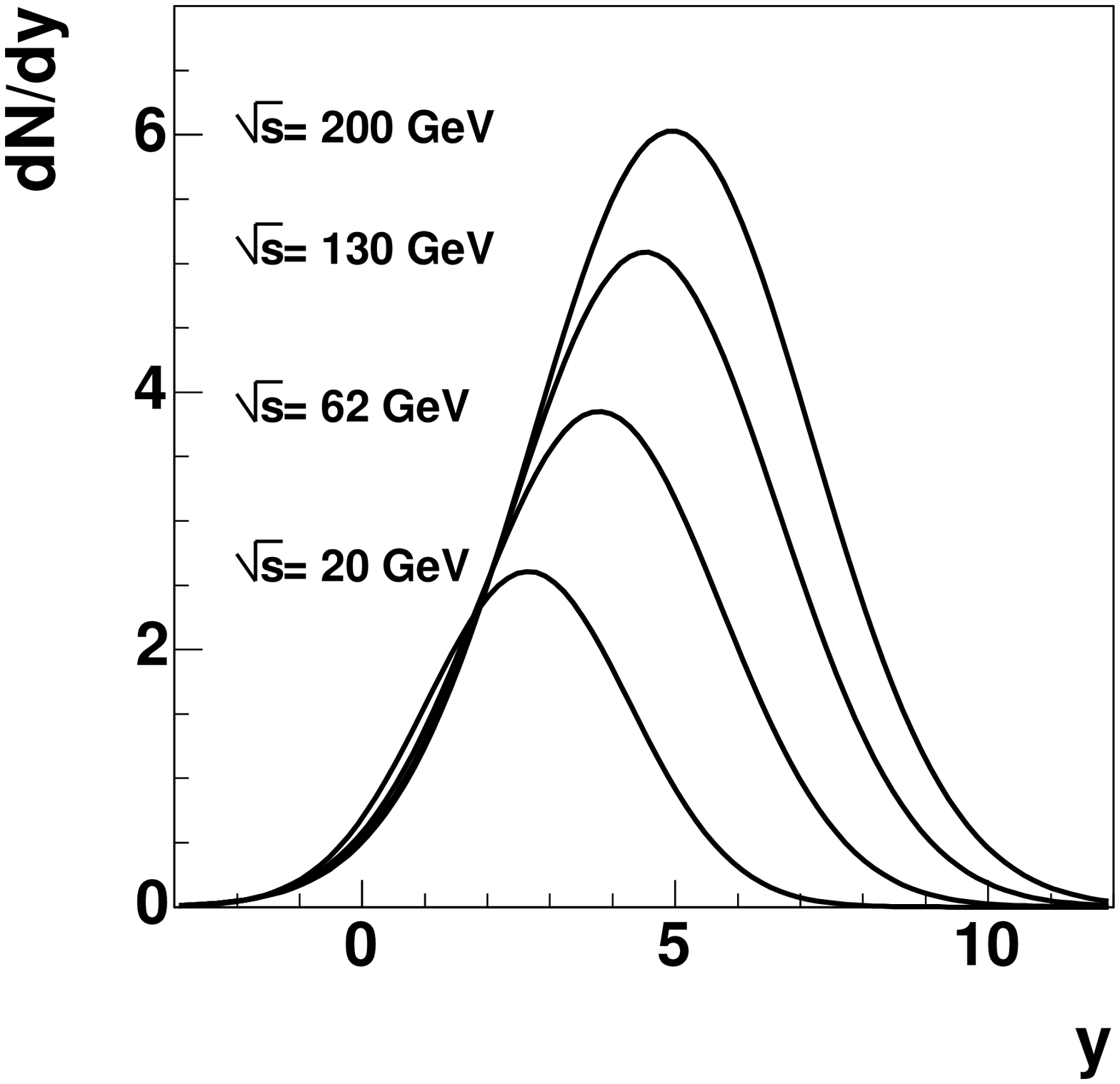}
\caption{
Results from the Landau formula, Equation (\ref{lanfull}),
in the rest frame of one of the projectiles.  Approximate
limiting fragmentation is observed.
\label{landau-limfrag}
}
\end{minipage}
\end{center}
\end{figure}

While the experimental status of Landau Gaussians vs. Feynman-Bjorken
plateaus \cite{Feynman:ej,Bjorken:1982qr} in dN/dy
was ambiguous throughout the 1970's \cite{Carruthers:ws,Carruthers:dw},
despite strong evidence for the applicability of Landau's formulas,
the situation in A+A collisions was clarified rather quickly
by the two large-acceptance RHIC experiments, PHOBOS and
BRAHMS.  PHOBOS quickly established that boost invariance was
violated over a large rapidity range by inclusive measurements
of $dN/d\eta$ over $|\eta|<5.4$ \cite{Back:2002wb}.  BRAHMS consolidated these
observations by finding that the rapidity distributions of 
pions in $|y|<3$ is Gaussian with a width parameter only 10\%
different from the Landau prediction \cite{Bearden:2004yx}.  
This led to a re-evaluation
of all of the existing heavy ion data, where it was found that
charged pion rapidity distributions all fell close to the Landau
trend \cite{Murray:2004gh,Roland:2004}.  

\section{Connections or Coincidences}

\begin{figure}[t]
\begin{center}
\begin{minipage}{60mm}
\includegraphics[width=6cm]{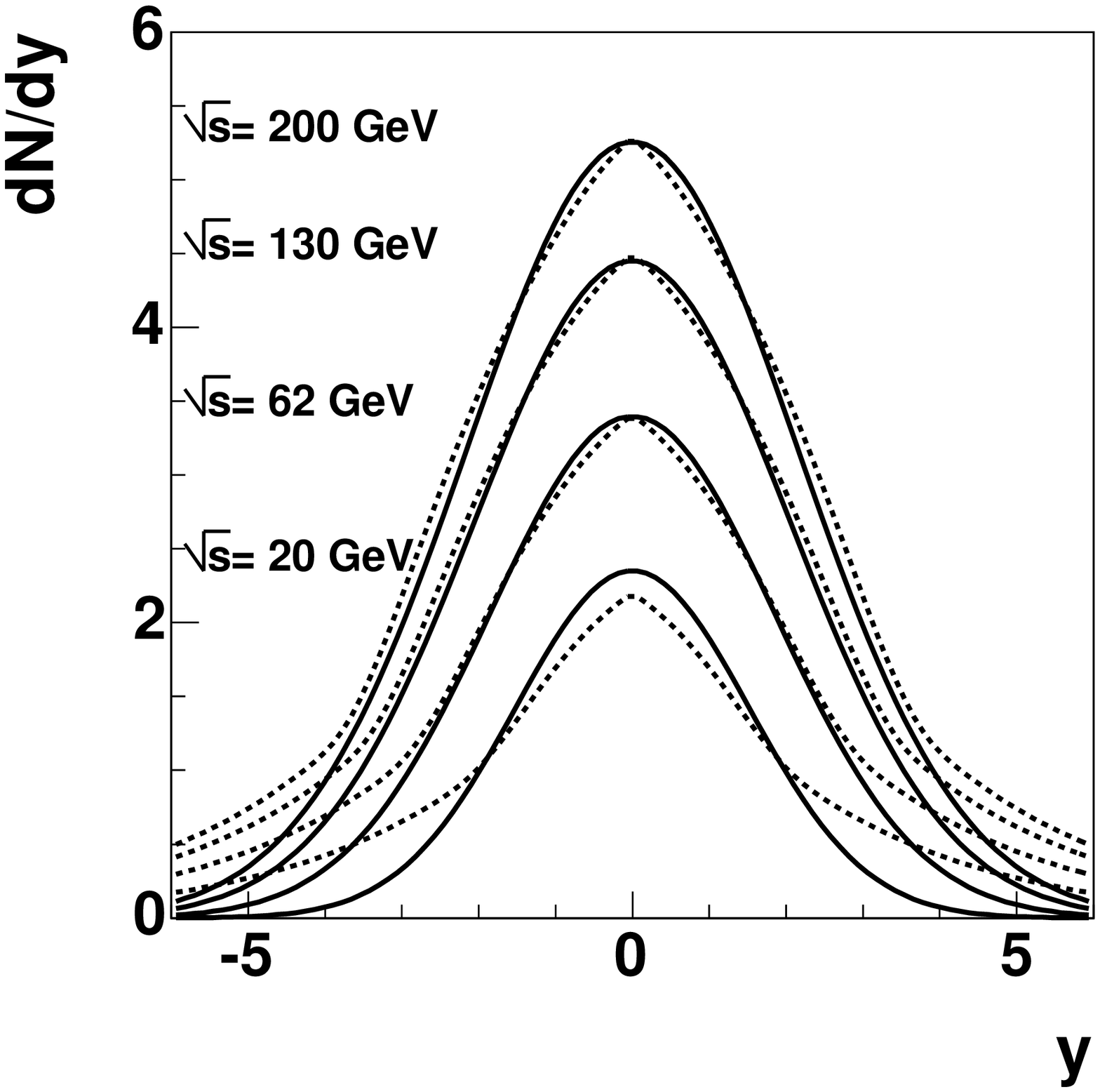}
\caption{
Comparison of the Landau result, Equation \ref{lanfull}, compared
with CGC results from Ref. \cite{Kharzeev:2001gp}.
The results are normalized at y=0 at 200 GeV.
\label{landau-cgc}
}
\end{minipage}
\hspace{\fill}
\begin{minipage}{60mm}
\includegraphics[width=6cm]{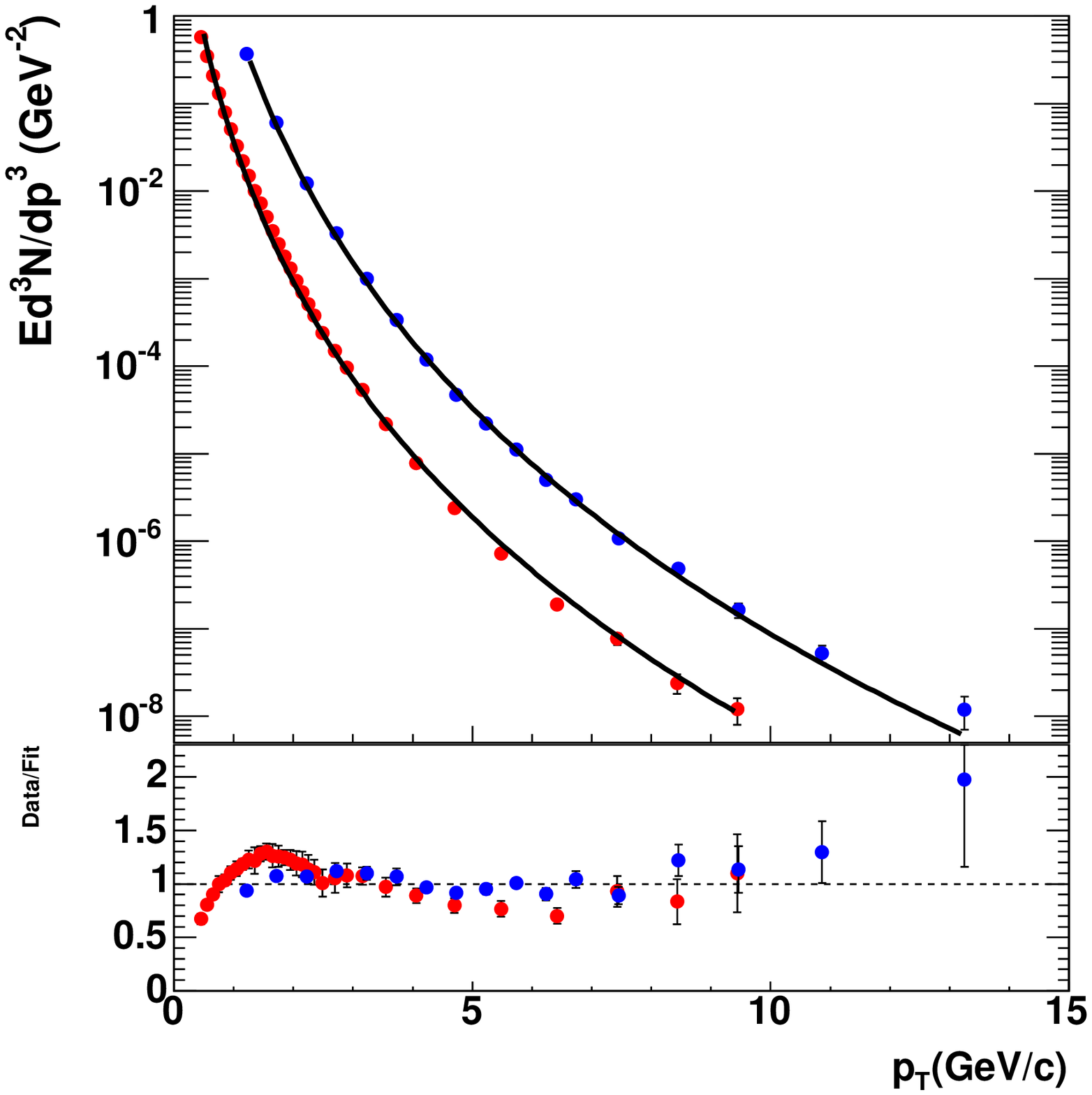}
\caption{
Top: STAR data on $Ed^3N/dp^3$ and PHENIX data on 
$Ed^3\sigma/dp^3$ fit to a Gaussian in $y_T$.
Bottom: Data divided by the fit.
\label{yt-gaussian}
}
\end{minipage}
\end{center}
\end{figure}

While the Landau results seem describe, and even predict,
several non-trivial results at RHIC, the formulas also
encode features that are usually attributed to QCD, or its various
approximations.

For example, measurements of $dN/d\eta$ in p+p and A+A collisions, 
boosted into the
rest frame of one of the projectiles, showed the phenomenon of
``limiting fragmentation'' \cite{Back:2002wb}.  
This is energy-independence of the particle
yields at a fixed rapidity distance from the projectile rapidity
\cite{Benecke:sh}.
One thing that has not typically been appreciated in the context 
of the Landau approach is that limiting fragmentation seems to arise
naturally from Equation (\ref{lanfull}):
Transforming this distribution into the rest frame of one of
the projectiles, $y^{\prime} = y-y_{beam}$, one finds that approximately:
\begin{equation}
\frac{dN}{dy^{\prime}}\propto
\frac{1}{\sqrt{L}}\exp(-\frac{y^{\prime 2}}{2L}-y^{\prime}).
\end{equation}
Which only varies weakly with $L$ for $y^{\prime}\sim 0$.
This seems too strong to be merely a coincidence, but it is not
clear why limiting fragmentation, which implies $x_F$ scaling,
arises naturally from the Landau formulas, in which $x_F$ does
not seem to be a preferred variable \cite{Carruthers:dw}.

Another interesting, and unexpected, connection can be made
between the Landau expressions and the results from the
CGC-based model of Kharzeev and Levin \cite{Kharzeev:2001gp}.  
Once one fixes the peak dN/dy of both models, and then varies the energy,
it appears that the width of dN/dy varies in a similar way
as a function of beam energy, as shown in Fig. \ref{landau-cgc}. 
It is interesting that the
exponent extracted from HERA data, $\lambda = 0.25-0.3$ is 
surprisingly similar to the power seen in the Landau multiplicity
formula.   However, it is not clear why this similarity occurs,
or how robust it is.

A final unexpected coincidence is seen in the transverse direction
near y=0.  Carruthers and Duong-van noticed that the $p_T$ distribution
of $\pi^0$'s in $p+p$ collisions
was well described out to $p_T=10$ GeV by a Gaussian
distribution in transverse rapidity 
$y_T = \frac{1}{2}\ln(\frac{m_T+p_T}{m_T-p_T})$
with $L\sim 0.51$ \cite{Minh:sg}.
While no derivation was given for this phenomenological 
description, which holds
over 10 orders of magnitude, an argument was made on a similar basis
as for the Gaussian in the longitudinal direction.  To see if this
function continues to work well at RHIC energies, fits have been
made to 
PHENIX $\pi^0$ data \cite{Adler:2003pb} and 
STAR inclusive charged data \cite{Adams:2003kv} from $p+p$ collisions,
which are shown in Fig. \ref{yt-gaussian}.  Reasonable agreement
is found with the STAR data with $L=0.56$, 
despite the combination of various particle
species, and excellent agreement is found with the PHENIX data
with $L=0.54$, up to $p_T = 11$ GeV.  

\section{Conclusions}
In conclusion, it is argued that the arguments made by Landau
and collaborators in the mid-1950's appear to have a surprising
relevance for understanding RHIC phenomena.  The results on
the total multiplicity and the shape of the rapidity distributions
hold in a robust way for a variety of systems, including 
$p+p$, $\epem$ and Au+Au, perhaps providing a natural way to
understand the apparent universality seen in the total number
of charged particles.  Several interesting connections are found
between the Landau results and limiting fragmentation, CGC
calculations, and even particle production at very high $p_T$.
Understanding these connections may provide deeper insight into
the strong interaction and the dynamical properties of 
strongly interacting systems.

\section*{Acknowledgments}
The author would like to thank the organizers for an enjoyable
and stimulating workshop - one that provided every possible opportunity
to combine work and leisure.  
Special thanks to 
Mark Baker,
Wit Busza, 
Dima Kharzeev, 
Jamie Nagle 
Gunther Roland, 
Gabor Veres
and
Bill Zajc
for illuminating discussions.

\vfill\eject
\end{document}